\documentclass[referee]{raa}           

\usepackage{graphicx,times}
\usepackage{natbib}
\usepackage{amssymb,amsmath}
\bibpunct{(}{)}{;}{a}{}{,}

\usepackage[pagebackref=true]{hyperref}

\begin{document}

   \title{Empirically Predicted Absolute Magnitudes for Red Clump Stars in Mephisto and CSST Filters}

 \volnopage{ {\bf 20XX} Vol.\ {\bf X} No. {\bf XX}, 000--000}
   \setcounter{page}{1}

   \author{Zheng Yu
   \inst{1}, Bing-Qiu Chen\inst{1}, Xiao-Wei Liu\inst{1}
   }

   \institute{$^1$South-Western Institute for Astronomy Research, Yunnan University, Chenggong District, Kunming 650091, P.\,R. China; {\it bchen@ynu.edu.cn; x.liu@ynu.edu.cn}\\
\vs \no
   {\small Received 20XX Month Day; accepted 20XX Month Day}
}

\abstract{Red clump (RC) stars are reliable standard candles for studying the structure and evolution of the Milky Way. In this study, we present empirical calibrations of RC absolute magnitudes in the Mephisto ($v, ~g, ~r, ~i$) and CSST ($g, ~r, ~i$) photometric systems using a high-purity sample of 25,059 RC stars cross-matched between APOGEE and Gaia DR3 XP spectra. Through synthetic photometry and polynomial fitting, we find that RC absolute magnitudes exhibit strong dependencies on effective temperature and metallicity, with the strongest variations observed in bluer bands and progressively decreasing towards redder wavelengths. In particular, the Mephisto $v$ band exhibits the highest sensitivity, with variations reaching up to 2.0\,mag across the metallicity range ($-$1.0\,dex $< $[Fe/H] $<$ 0.5\,dex) and the temperature range (4500--5200\,K). The calibrations achieve high precision for all bands, enabling accurate determination of RC absolute magnitudes and distances. Furthermore, we evaluate the metallicity estimation capabilities of both systems using a Random Forest-based method, achieving a precision of 0.12\,dex for Mephisto and 0.14\,dex for CSST under typical photometric uncertainties ($\leq$ 0.01\,mag). These results provide robust tools for distance and metallicity determinations, supporting future Galactic structure studies with Mephisto and CSST data.
}

   \authorrunning{Zheng Yu et al. }            
   \titlerunning{Absolute Magnitudes for Red Clump Stars}  
   \maketitle

%
\section{Introduction}           
\label{sect:intro}

Red clump (RC) stars are low-mass, core-helium-burning giants with relatively stable luminosities, making them reliable standard candles for distance measurements and Galactic structure studies \citep{cannon1970red, paczynski1998galactocentric}. Their distinct concentration in color-magnitude diagrams (CMDs) across various stellar populations allows for their widespread use in studying the structure and evolution of the Milky Way \citep{girardi2016red}. RC stars have been extensively employed in mapping the three-dimensional structure and kinematics of the Galaxy \citep[e.g.,][]{stanek1997modeling, lopez2002old, wegg2013mapping, yu2021mapping, yu2025stellar}, measuring interstellar extinction \citep{nataf2013reddening, de2014probing, wang2019optical}, and tracing chemical gradients in the Galactic disk \citep{bovy2014apogee, huang2015metallicity, bovy2016stellar}.

Despite their advantages, RC stars are not perfect standard candles. Their absolute magnitudes depend on stellar parameters such as metallicity and age—an effect commonly referred to as 'population effects' \citep{girardi1999secondary, salaris2002population, pietrzynski2010araucaria, girardi2016red, chen2017absolute}. Ignoring these variations can introduce significant systematic errors in distance estimates, particularly when comparing regions with different star formation histories and chemical evolution patterns \citep{girardi2001population, ruiz2018empirical}. Therefore, accurately characterizing these dependencies is crucial for maximizing the usefulness of RC stars in Galactic studies.

Considerable effort has been made to empirically calibrate RC absolute magnitudes across various photometric systems. Early calibrations were primarily based on Hipparcos data \citep{paczynski1998galactocentric, alves2000k, groenewegen2008red, laney2012new}, while more recent studies have leveraged high-precision astrometry from Gaia \citep{chen2017absolute, hawkins2017red, ruiz2018empirical}. \citet{ruiz2018empirical} provided a comprehensive calibration for RC stars in Gaia DR1, demonstrating clear dependencies of absolute magnitudes on color, effective temperature, and metallicity. Similarly, \citet{huang2020mapping} combined Gaia DR2 parallaxes with LAMOST spectroscopy to refine RC absolute magnitudes in the near-infrared $K_{S}$ band, achieving distance precision as high as 5–10\%, surpassing Gaia’s own parallax-based measurements at distances beyond $\sim$4\,kpc.

The upcoming optical surveys conducted by the China Space Station Telescope (CSST; \citealt{zhan2011consideration, cao2018testing}) and the Multi-channel Photometric Survey Telescope (Mephisto; \citealt{Yuan2020, Lei2021, Lei2022}) will provide unprecedentedly large and precise photometric datasets. CSST, a planned 2-meter-class space telescope scheduled for launch in 2026, will conduct deep imaging in seven broad bands ($NUV$, $u$, $g$, $r$, $i$, $z$, $y$) over approximately 17,500\,deg$^2$. Mephisto, located at Lijiang Observatory, is uniquely capable of simultaneous multi-band imaging in six filters ($u$, $v$, $g$, $r$, $i$, $z$) and will survey about 20,000\,deg$^2$ of the northern sky. These datasets will significantly enhance our ability to study Galactic structure and stellar populations.

The primary goal of this study is to accurately determine the absolute magnitudes of RC stars in the optical bands of CSST and Mephisto and to quantify their dependencies on stellar parameters. This calibration will be essential for precisely flux-calibrating future CSST and Mephisto surveys, enabling accurate distance estimates and atmospheric parameter determinations for RC stars. These improvements will further our understanding of the Milky Way’s structure and evolution. This paper is structured as follows. In Section~\ref{sec:2}, we describe the data and methods. Section~\ref{sec:3} presents our empirical calibrations of absolute magnitudes in Mephisto and CSST bands and explores their dependencies on stellar parameters. In Section \ref{sec:4}, we discuss our results in the context of existing theoretical predictions and empirical studies, highlighting improvements and implications for future Galactic research. Finally, in Section \ref{sec:5}, we summarize our main conclusions.

\section{Data and Method} \label{sec:2}

This study aims to establish accurate calibrations of RC absolute magnitudes in the Mephisto ($v$, $g$, $r$, $i$) and CSST ($g$, $r$, $i$) photometric bands using Gaia XP spectra. These bands were chosen because they fall within the wavelength range of Gaia XP spectra (330--1050\,nm; \citealt{de2023gaia}), whereas other bands (such as Mephisto's $u$, $z$ and CSST's $NUV$, $u$, $z$, $y$) extend beyond this range and will be analyzed in future work. 

We construct a high-purity RC sample based on the selection by \citet{ting2018large}, who identified RC stars using data from the Apache Point Observatory Galactic Evolution Experiment (APOGEE; \citealt{majewski2017apache}). These stars are then cross-matched with Gaia DR3 XP ultra-low-resolution spectra \citep{vallenari2023astronomy, huang2024comprehensive}, allowing us to derive synthetic photometry and obtain precise absolute magnitudes. We systematically quantify the dependencies of absolute magnitudes on effective temperature, metallicity, and surface gravity. 

\subsection{RC Sample} \label{sec:2.1}

A clean and well-characterized RC sample is essential for reliable absolute magnitude calibrations. We adopt the primary RC sample identified by \citet{ting2018large} from APOGEE data. Using a neural network trained on Kepler asteroseismic observations, they identified primary RC stars based on their characteristic period spacing $(\Delta P > 250\,\text{s})$, which effectively separates them from RGB and secondary RC stars, even when their atmospheric parameters are similar. Cross-validation tests confirmed a low contamination rate of 2--3\% in the RC classification. 

The atmospheric parameters of these stars were obtained from high-resolution APOGEE spectra ($R \sim 22,500$) using the APOGEE Stellar Parameter and Chemical Abundances Pipeline (ASPCAP; \citealt{perez2016aspcap}). Based on the use of ASPCAP as reported in \citet{accetta2022seventeenth} for DR17, the stellar effective temperatures ($T_{\text{eff}}$), surface gravities (log\,$g$), and metallicities ([Fe/H]) are precise to approximately 2\%, 0.1\,dex, and 0.05\,dex, respectively.

We cross-match this RC catalog with Gaia DR3 XP spectra using a 1-arcsecond matching radius. Gaia XP spectra provide ultra-low-resolution (${\lambda}/{\Delta \lambda} \sim 50 $) spectroscopic data covering 336--1020\,nm for approximately 220 million stars with $G < 18$\,mag. For our analysis, we use the recalibrated XP spectra from \citet{huang2024comprehensive}, who corrected systematic offsets in color, magnitude, reddening, and wavelength by comparing Gaia DR3 XP spectra with well-calibrated spectral standards from CALSPEC \citep{bohlin2022update}, NGSL \citep{koleva2012stellar}, and LAMOST \citep{cui2012large}. The initial cross-matching results in a sample of 33,912 RC stars. 

\subsection{Extinction Correction} \label{sec:2.2}

Accurate extinction corrections are necessary to obtain reliable absolute magnitudes. We determine interstellar extinction for each RC star based on the extinction values from \citet{zhang2023parameters}, who derived stellar parameters and extinction for 220 million stars using a data-driven model incorporating Gaia XP spectra, 2MASS \citep{skrutskie2006two}, and unWISE \citep{schlafly2019unwise} photometry. In their model, the extinction at wavelength $\lambda$ for a given star $k$ is expressed as,
\begin{equation}
A_{\lambda,k} = E_k \times R_{\lambda},
\end{equation}
where $E_k$ represents the overall extinction strength for the star, and $R_{\lambda}$ describes the wavelength dependence of extinction. For each target RC star, we compute a weighted mean extinction value using nearby stars within a 10\,pc radius, given by,
\begin{equation}
\bar{E} = \frac{\sum (E_i \times w_i)}{\sum w_i},
\label{eq:1}
\end{equation}
where $E_i$ is the extinction of the $i$-th nearby star, and $w_i$ is its weight. Following \citet{chen2024constructing}, the weight is given by,
\begin{equation}
w_i = \frac{1 - 0.75 \left(\frac{d_i}{d_{\text{max}}} \right)^2}{\sigma_{E_i}^2 + 10^{-4}},
\label{eq:2}
\end{equation}
where $d_i$ is the distance between the target and the nearby star, $d_{\text{max}}$ is the maximum distance among all nearby stars, and $\sigma_{E_i}$ is the uncertainty in extinction. Using the extinction curve $R_{\lambda}$ from \citet{zhang2023parameters}, we compute the extinction at each wavelength, as,
\begin{equation}
A_{\lambda} = \bar{E} \times R_{\lambda}.
\label{eq:3}
\end{equation}
These extinction values are then used to correct the Gaia XP spectra fluxes at each wavelength.

\begin{figure}
    \centering
    \includegraphics[width=1\linewidth]{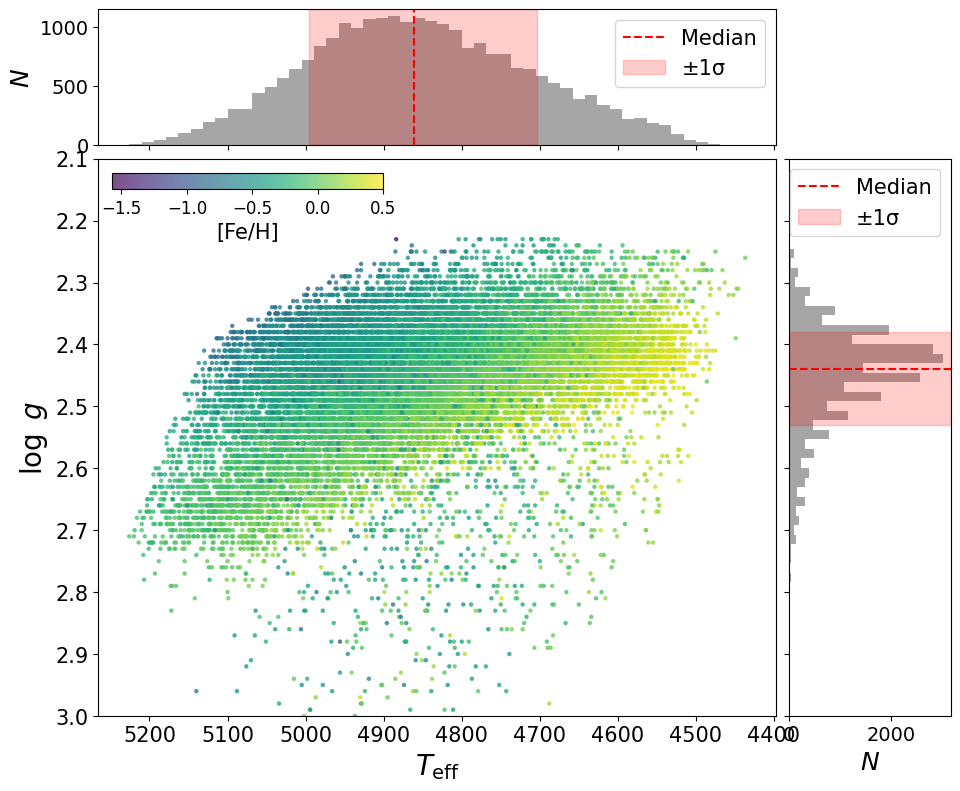}
    \caption{Distribution of our RC sample in the Kiel diagram ($T_{\text{eff}}$ vs. log\,$g$). The points are color-coded by [Fe/H]. The top panel shows the histogram of $T_{\text{eff}}$, while the right panel presents the log\,$g$ distribution. Red dashed lines indicate median values, and pink shaded regions represent the $\pm 1\sigma$ ranges.}
    \label{Figure1}
\end{figure}

To further obtain the absolute magnitudes, we adopt distances from \citet{zhang2023parameters}.  We restrict our sample to stars with relative distance errors below 10\%. After applying this criterion, our final sample consists of 25,059 RC stars. The atmospheric parameters of our final sample exhibit well-defined distributions consistent with RC stars, as illustrated in Fig.~\ref{Figure1}. The effective temperature ($T_{\text{eff}}$) ranges from 4500 to 5200\,K, with a median of 4860\,K. The surface gravity (log\,$g$) is tightly distributed between 2.2 and 2.8\,dex, peaking at 2.4\,dex, which aligns with expectations for RC stars. The metallicity ([Fe/H]) spans $-$1.5 to 0.5\,dex, with most stars clustering around solar metallicity and relatively few having [Fe/H] $< -1.0$\,dex. 

\subsection{Synthetic Photometry} \label{sec:2.3}

To derive photometry from Gaia XP spectra, we perform synthetic photometry using the transmission curves of the Mephisto ($v$, $g$, $r$, $i$) and CSST ($g$, $r$, $i$) filter systems. For each filter, the intrinsic apparent magnitude $m_0$ is calculated as,
\begin{equation}
m_0 = -2.5 \log_{10} \frac{\int f_{\lambda,0} T_\lambda \lambda\, d\lambda}{\int f_{\text{zero}(AB)} T_\lambda \lambda \, d\lambda},
\label{eq:6}
\end{equation}
where $f_{\lambda,0}$ is the extinction-corrected flux from Section~\ref{sec:2.2}, $T_\lambda$ is the filter transmission function, and $f_{\text{zero}(AB)}$ is the AB magnitude zero-point flux. The uncertainty in $m_0$ is determined by propagating the uncertainties in the extinction-corrected flux:
\begin{equation}
\sigma_{m_0} = \frac{2.5}{\ln(10)} \times \left( \frac{\sigma_F}{F} \right),
\label{eq:7}
\end{equation}
where $F$ is the total flux and $\sigma_F$ is the weighted total flux uncertainty.

\begin{figure}
    \centering
    \includegraphics[width=1\linewidth]{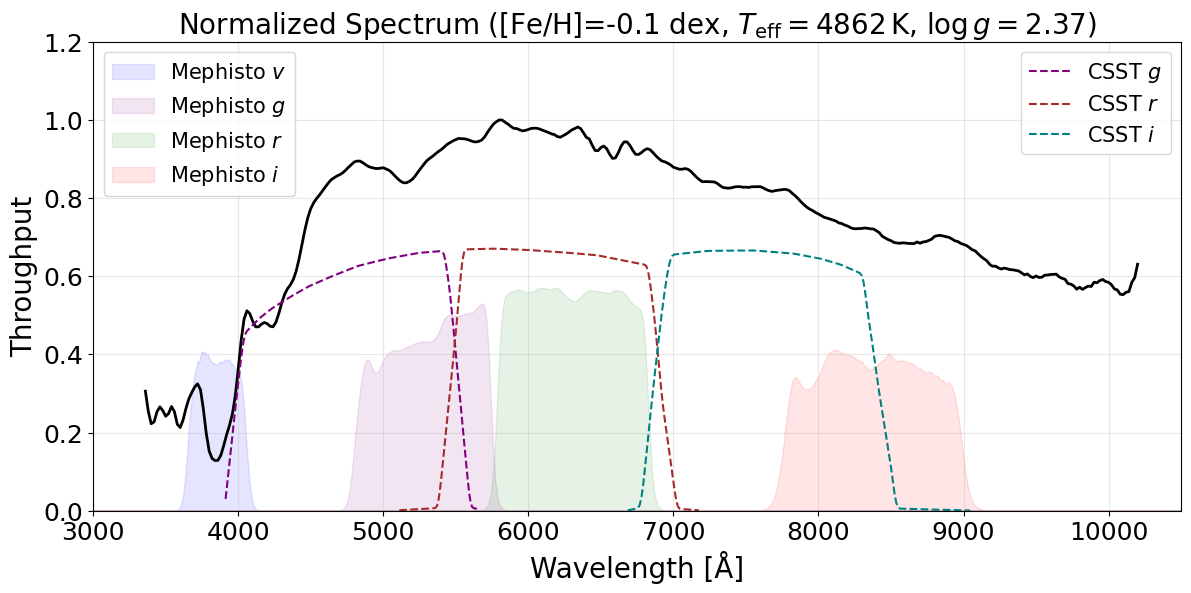}
    \caption{The normalized Gaia XP spectrum of an example RC star with parameters [Fe/H] = --0.1\,dex, $T_{\mathrm{eff}}$ = 4862\,K, and $\log g$ = 2.37\,dex, shown in black. The transimission curves of Mephisto filters ($v$, $g$, $r$, $i$) are represented by the filled regions, while the CSST filter transimission curves ($g$, $r$, $i$) are indicated by dashed lines.}
    \label{Figure_transmission}
\end{figure}

Fig.~\ref{Figure_transmission} illustrates the normalized Gaia XP spectrum of a representative RC star ([Fe/H] = $-$0.1\,dex, $T_{\mathrm{eff}} = 4862$\,K, $\log g = 2.37$\,dex) alongside the transmission curves of the Mephisto and CSST filters. The filled regions represent the Mephisto ($v$, $g$, $r$, $i$) filter transmission curves, while the dashed lines correspond to the CSST ($g$, $r$, $i$) filters. The CSST transmission curves are based on the detailed parameters described in \citet{cao2018testing}, while the Mephisto transmission curves are adopted from \citet{chen2024early}.

Using the distance estimates from \citet{zhang2023parameters}, we calculate absolute magnitudes by,
\begin{equation}
M = m_0 - 5 \log_{10}(d) + 5.
\label{eq:10}
\end{equation}

The uncertainty in the absolute magnitude, $\sigma_M$, accounts for both the intrinsic photometric uncertainty, $\sigma_{m_0}$, and the distance uncertainty, $\sigma_d$:
\begin{equation}
\sigma_M = \sqrt{\sigma_{m_0}^2 + \left( \frac{5 \sigma_d}{d \ln 10} \right)^2}.
\label{eq:11}
\end{equation}

\section{Results}\label{sec:3}

\begin{figure}
    \centering
    \includegraphics[width=1\linewidth]{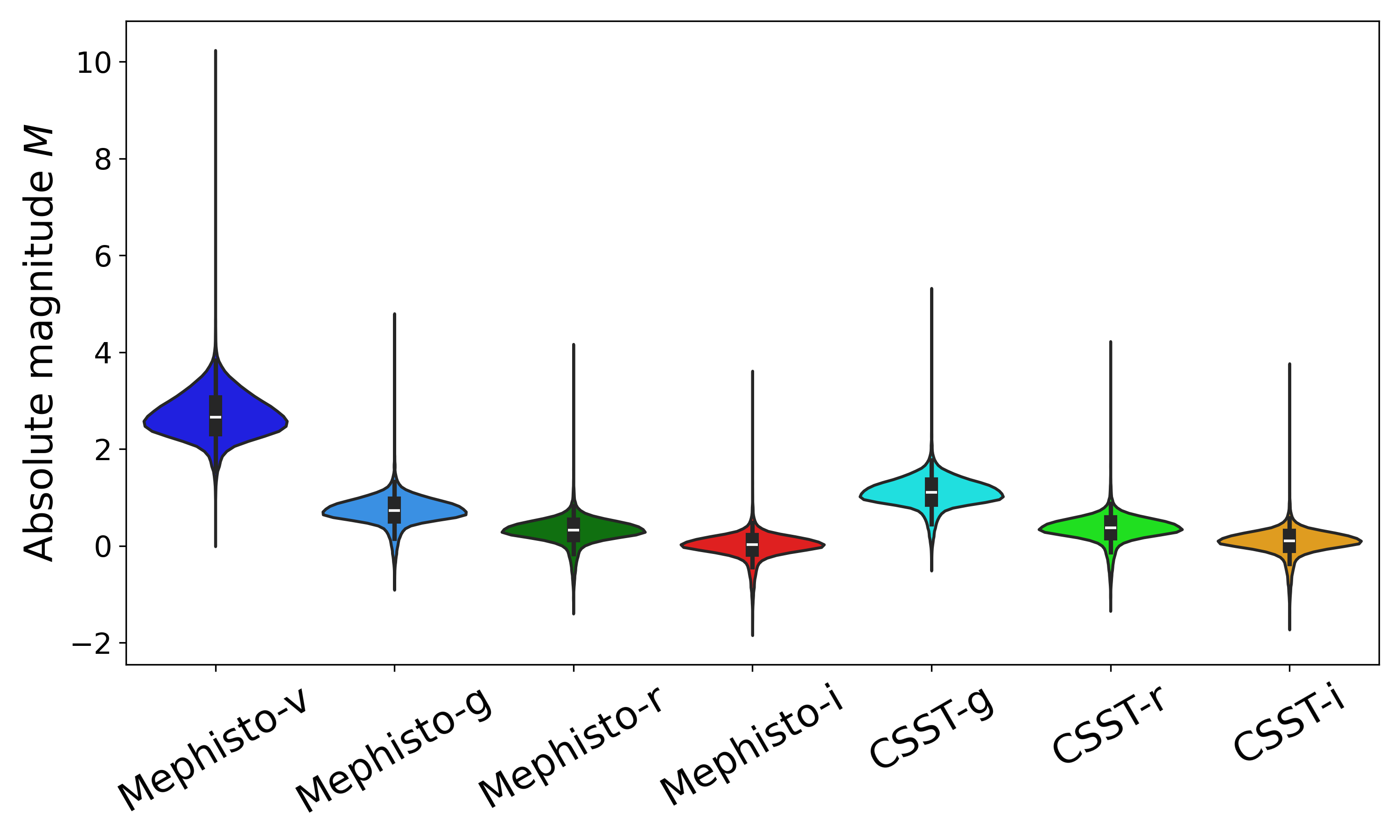}
    \caption{Distribution of absolute magnitudes ($M$) for the RC star sample. The width of the violin plots represents the density of data points at each magnitude value. The internal box plots indicate the median (white lines), interquartile range (thick black bars), and whiskers extending to 1.5 times the interquartile range.
    }
    \label{Figure3}
\end{figure}

\begin{table*}
\centering
\caption{Median and dispersion of absolute magnitudes ($M$) in different photometric bands.}
\label{tab:1}
\begin{tabular}{cccccccr}
\hline
\hline
& Mephisto  & Mephisto  & Mephisto    & Mephisto  & CSST   & CSST   & CSST  \\
&  $M_v$ &  $M_g$ &  $M_r$ &  $M_i$ &  $M_g$ &  $M_r$ &  $M_i$ \\
\hline
Median (mag) & 2.66 & 0.74 & 0.33 & 0.02 & 1.10 & 0.37 & 0.10 \\
Dispersion (mag) & 0.44 & 0.27 & 0.25 & 0.24 & 0.29 & 0.25 & 0.25 \\
\hline
\end{tabular}
\end{table*}

\begin{figure}
    \centering
    \includegraphics[width=1\linewidth]{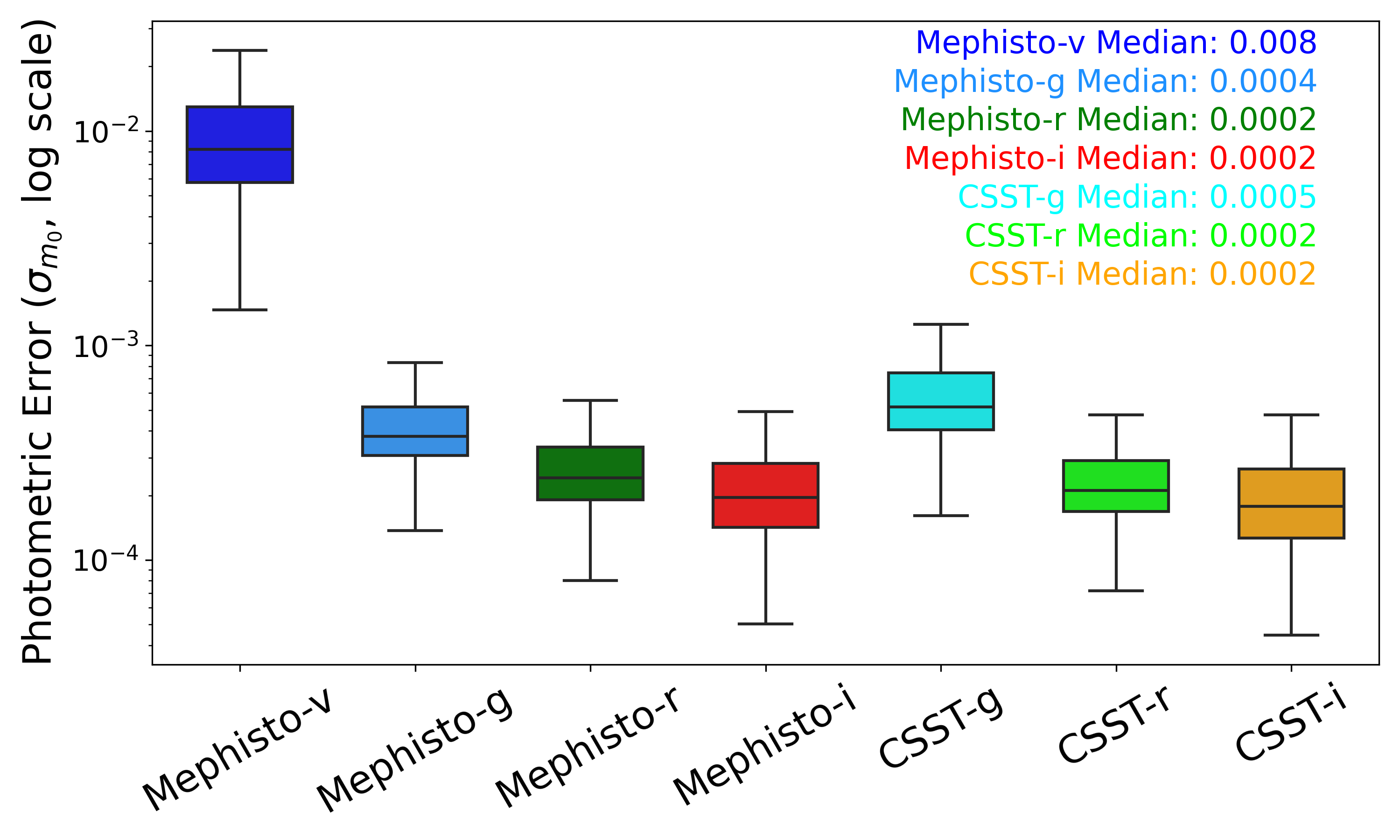}
    \caption{Photometric uncertainties in intrinsic apparent magnitudes, $\sigma_{ m_0}$, for the RC stars across the Mephisto ($v$, $g$, $r$, $i$) and CSST ($g$, $r$, $i$) bands, displayed on a logarithmic scale. The box plots represent the quartiles of the error distribution, with the central line indicating the median, box edges marking the first and third quartiles, and whiskers extending to 1.5 times the interquartile range.
    }
    \label{Figure2}
\end{figure}

We computed the empirical absolute magnitudes for all RC stars in our sample. Fig.~\ref{Figure3} presents the distributions of absolute magnitudes in the Mephisto ($v$, $g$, $r$, $i$) and CSST ($g$, $r$, $i$) photometric systems. The violin plots illustrate the overall distributions, with internal box plots highlighting the medians and interquartile ranges. The corresponding statistical summaries are provided in Table~\ref{tab:1}.

In both photometric systems, the absolute magnitudes systematically decrease from blue to red bands. In the Mephisto system, the median absolute magnitude decreases from 2.66\,mag in the $v$ band to 0.02\,mag in the $i$ band. Similarly, in the CSST system, the median values range from 1.10\,mag in the $g$ band to 0.10\,mag in the $i$ band. The dispersion of absolute magnitudes also decreases across the bands. For the Mephisto system, the dispersion reduces from 0.44\,mag in the $v$ band to 0.24\,mag in the $i$ band, while in the CSST system, it declines from 0.29\,mag in the $g$ band to 0.25\,mag in the $i$ band. The larger dispersions observed in the blue bands suggest that absolute magnitudes are more strongly influenced by population effects in these wavelengths.

\subsection{Absolute magnitude uncertainties} 

In addition to population effects, the dispersion in absolute magnitude is also affected by measurement uncertainties. The uncertainty in absolute magnitude, $\sigma_M$, arises from two main sources: the intrinsic photometric uncertainty, $\sigma_{m_0}$, and the uncertainty in distance, $\sigma_d$. We first examine $\sigma_{m_0}$, which is illustrated in Fig.~\ref{Figure2} using box plots on a logarithmic scale. The photometric uncertainties are generally small, with median values ranging from 0.0002 to 0.008\,mag. The Mephisto $v$ band exhibits the highest uncertainty (median = 0.008\,mag), likely due to greater flux uncertainties in the blue region of the Gaia XP spectra. The remaining bands demonstrate much higher precision, with median uncertainties of 0.0004\,mag in the Mephisto $g$ band, 0.0005\,mag in the CSST $g$ band, and approximately 0.0002\,mag in the other bands. These results highlight the high photometric accuracy of our measurements, which is crucial for determining precise stellar parameters, including metallicity, as discussed in Section~\ref{sec:3.3}.

For our sample, the uncertainty in absolute magnitude, $\sigma_M$, is primarily determined by the distance uncertainty, $\sigma_d$, while the contribution from intrinsic photometric uncertainty, $\sigma_{m_0}$, is negligible. The median absolute magnitude uncertainty across all bands is approximately 0.07\,mag. This consistency across different bands indicates that the dominant source of error in $M$ stems from relative distance uncertainties ($\leq$ 10\%), which contribute significantly more than photometric errors ($<$ 0.01\,mag). Overall, the computed absolute magnitude uncertainty (0.07\,mag) is small compared to the observed dispersion in absolute magnitudes (0.24--0.44\,mag), suggesting that the dispersion is primarily driven by population effects rather than measurement errors.

\subsection{Dependencies on Stellar Parameters} \label{sec:3.2}

\begin{table*}
  \centering
  \small
  \caption{Polynomial fitting coefficients and RMSE values for absolute magnitude fitting.}
  \label{tab:2}
   \begin{tabular}{lccccccc}
\hline
\hline
Band         & $\beta_0$ & $\beta_1$ & $\beta_2$ & $\beta_3$ & $\beta_4$ & $\beta_5$  & RMSE (mag) \\
\hline
Mephisto $M_v$ & --76.789  & --4.767  & 33.766   & 0.115   & 5.191  & --12.009     & 0.26   \\
Mephisto $M_g$ & --30.127  & --1.999  & 18.666   & 0.176   & 2.039  & --7.012   & 0.24   \\
Mephisto $M_r$ & --6.664   & --1.157  & 15.874   & 0.195   & 1.152  & 6.201     & 0.24   \\
Mephisto $M_i$ & --1.338   & --0.223  & 12.041   & 0.265   & 0.266  & --4.809   & 0.24   \\
CSST $M_g$     & 150.556   & --2.682  & 23.658   & 0.158   & 2.762  & --9.044   & 0.24   \\
CSST $M_r$     & --10.689  & --1.222  & 16.442   & 0.192   & 1.222  & --6.427   & 0.24   \\
CSST $M_i$     & 11.446    & --0.555  & 13.583   & 0.247   & 0.588  & --5.403   & 0.24    \\
\hline
\end{tabular}
\end{table*}

\begin{figure*}
    \centering
    \includegraphics[width=\textwidth]{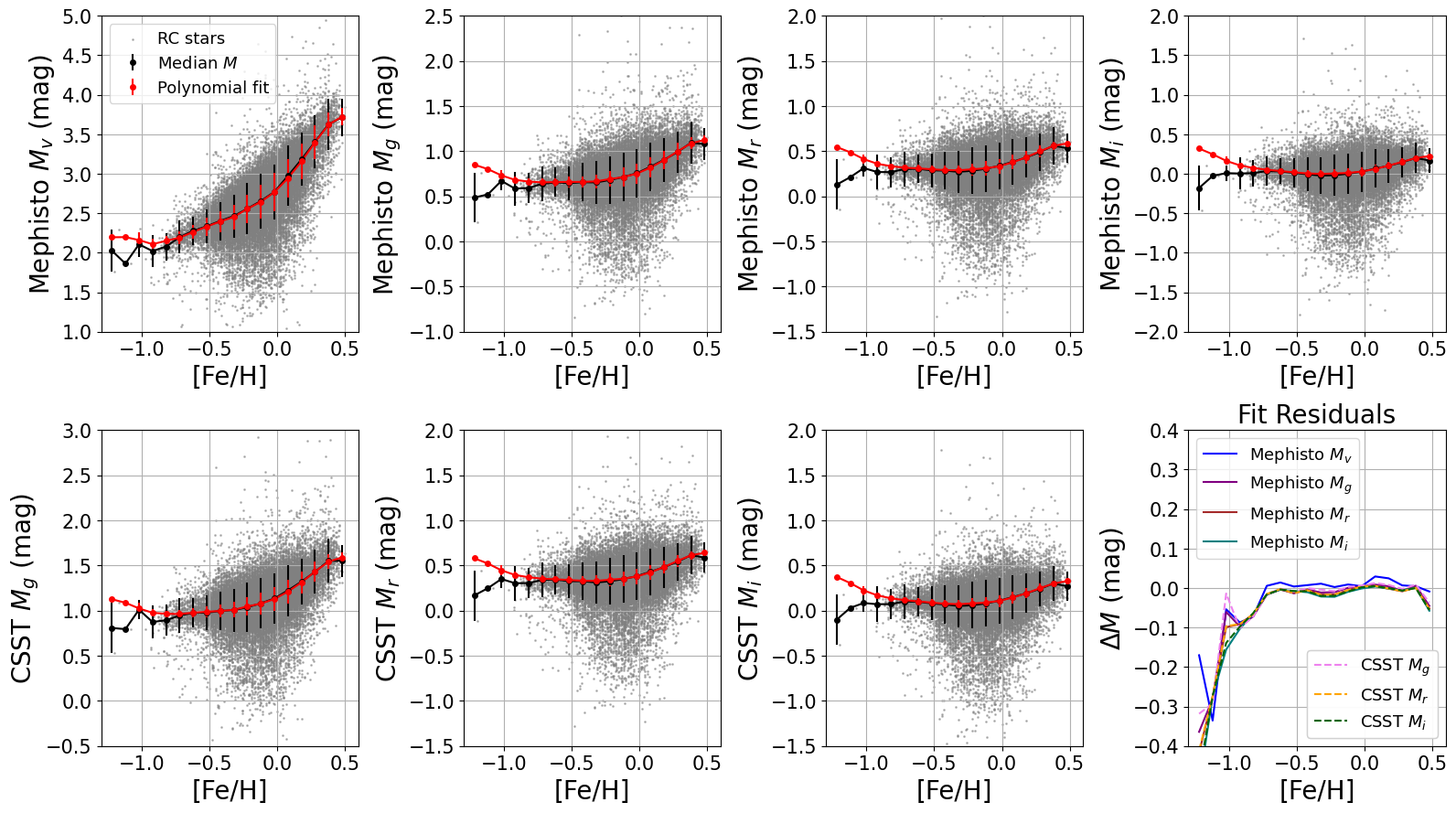}
    \caption{Relationship between absolute magnitudes and metallicity [Fe/H] for RC stars in the Mephisto and CSST photometric systems. Gray dots represent individual RC stars, while black dots with error bars indicate median values and their 1-$\sigma$ dispersions in each [Fe/H] bin. The red curves denote the best-fit polynomials described in Equation~\ref{eq:12}. The upper panels show Mephisto bands ($M_v$, $M_g$, $M_r$, $M_i$), while the lower panels show CSST bands ($M_g$, $M_r$, $M_i$). The rightmost panel displays residuals ($\Delta M$) between the polynomial fits and binned medians, with solid and dashed lines representing Mephisto and CSST bands, respectively. 
    }
    \label{figure4}
\end{figure*}

\begin{figure*}
    \centering
    \includegraphics[width=\textwidth]{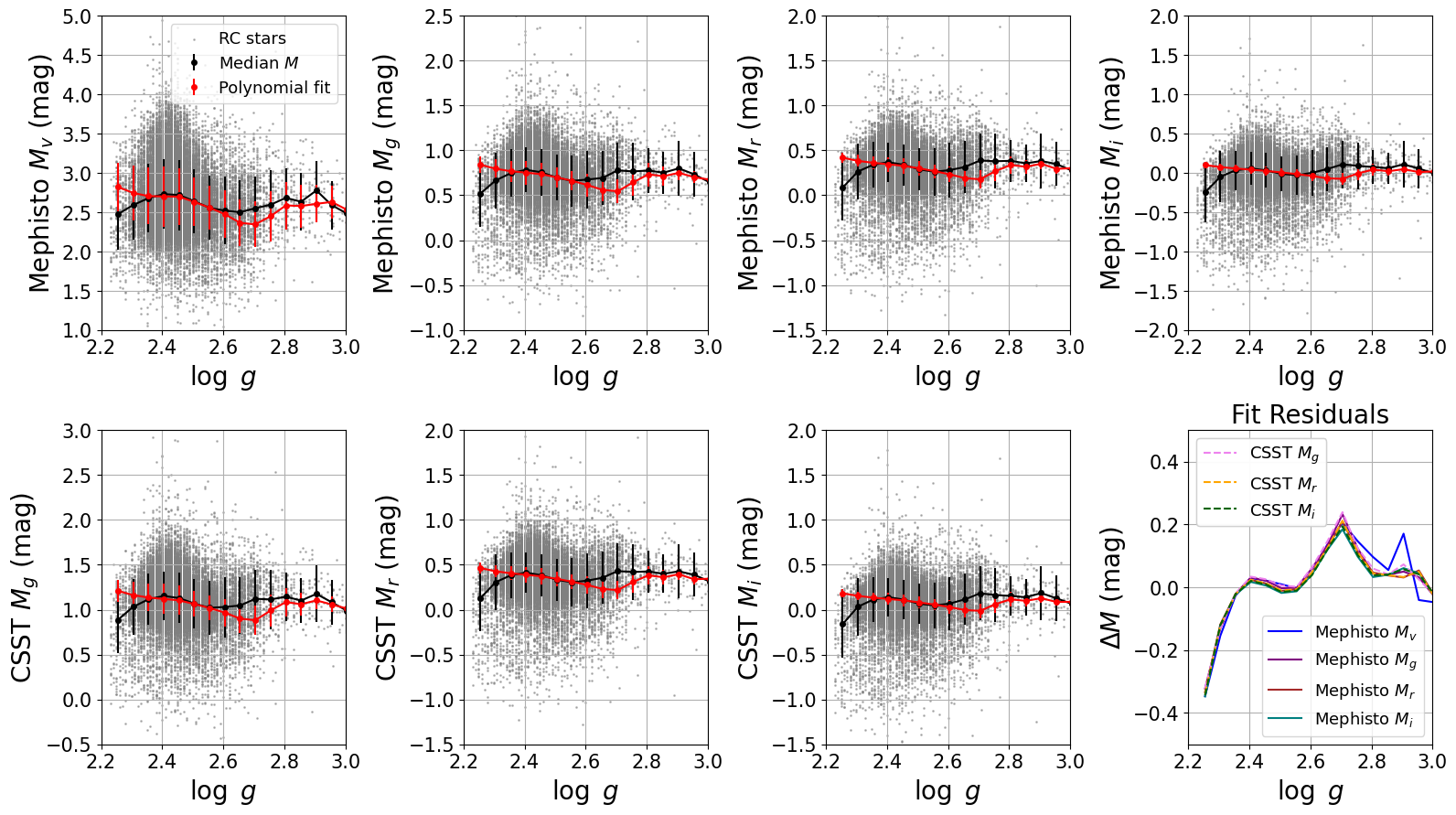}
    \caption{Similar to Fig.~\ref{figure4}, but showing the relationship between absolute magnitudes and surface gravity log\,$g$.}
    \label{figure5}
\end{figure*}

\begin{figure*}
    \centering
    \includegraphics[width=\textwidth]{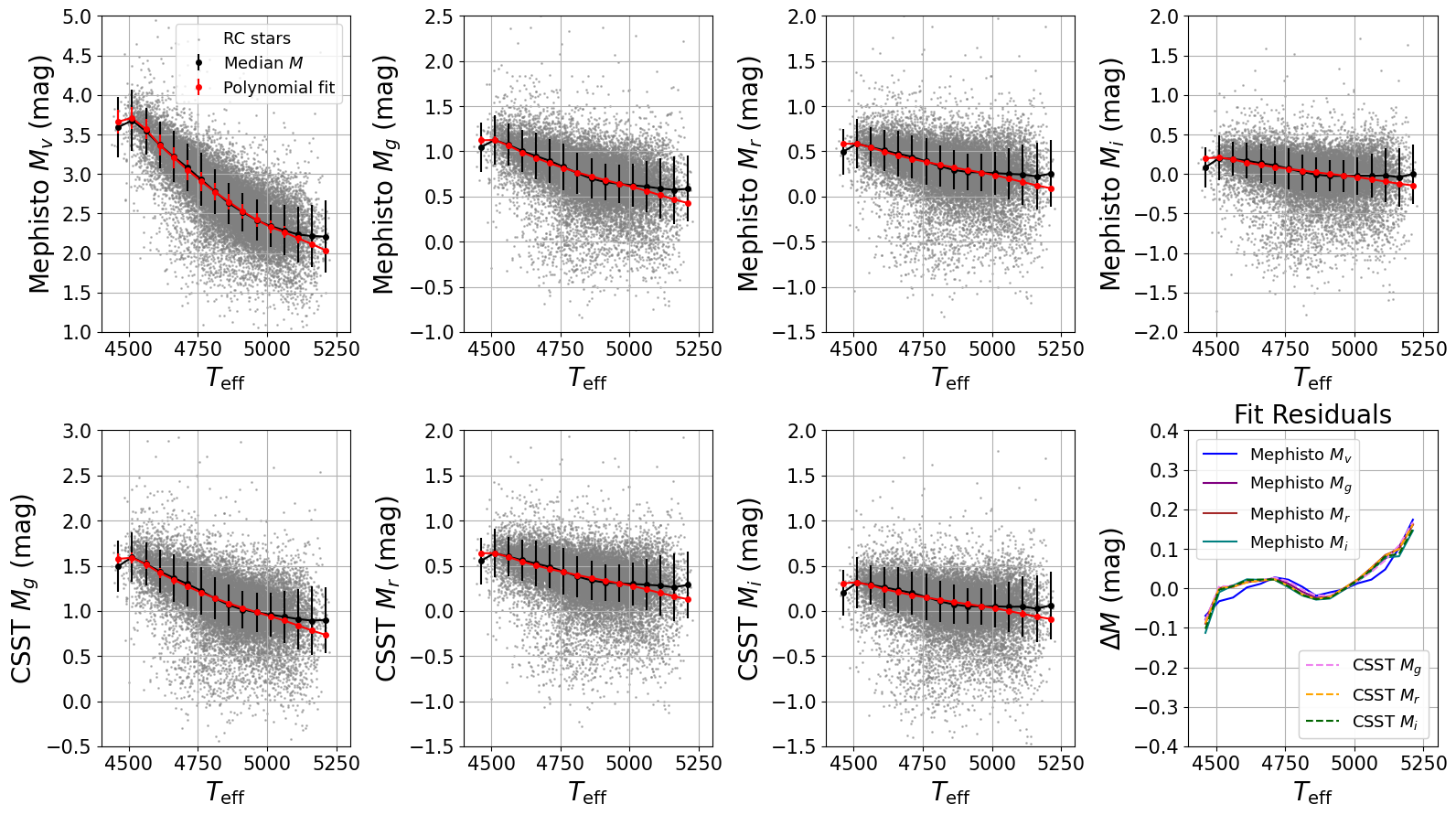}
    \caption{Similar to Fig.~\ref{figure4}, but showing the relationship between absolute magnitudes and effective temperature $T_{\rm eff}$.}
    \label{figure6}
\end{figure*}

To examine how absolute magnitudes of RC stars depend on stellar parameters, Figs.~\ref{figure4}, \ref{figure5} and \ref{figure6} show the relationships between absolute magnitude $M$ and metallicity ([Fe/H]), surface gravity (log\,$g$), and effective temperature ($T_{\rm eff}$) for both the Mephisto ($v$, $g$, $r$, $i$) and CSST ($g$, $r$, $i$) photometric systems.

Overall, absolute magnitudes tend to increase (i.e., stars become fainter) with higher [Fe/H]. This effect is strongest in the Mephisto $v$ band, where magnitudes vary by approximately 2.0\,mag over the full metallicity range ($-$1.0\,dex $<$ [Fe/H] $<$ 0.5\,dex). The sensitivity decreases toward longer wavelengths, with the $i$ band showing a variation of only about 0.5\,mag. As expected from the clustering of RC stars in the $T_{\text{eff}}$--log\,$g$ plane (see Fig.~\ref{Figure1}), our sample exhibits a narrow log\,$g$ distribution (primarily between 2.2\,dex and 2.8\,dex), characteristic of RC stars. Consequently, no significant trend is observed between absolute magnitude and log\,$g$. However, a clear negative correlation is evident between absolute magnitude and $T_{\rm eff}$ in all bands. The Mephisto $v$ band again shows the strongest dependence, with magnitudes varying by approximately 2.0\,mag over the temperature range (4500\,K $<$ $T_{\rm eff}$ $<$ 5200\,K). This sensitivity weakens toward redder bands, with the $i$ band showing a variation of only about 0.5\,mag.

To model these dependencies, we performed a second-order polynomial fit using weighted linear regression. The effective temperature was normalized as $T_{\text{eff, norm}} = T_{\text{eff}}/5000$\,K to improve numerical stability. The dataset was randomly split into training (80\%) and validation (20\%) subsets, with fitting weights assigned as the inverse square of absolute magnitude uncertainties to account for measurement errors. The polynomial fitting function is given by:
\begin{align}
M = &\ \beta_0 + \beta_1 x + \beta_2 y + \beta_4 x^2 + \beta_5 xy + \beta_7 y^2 
\label{eq:12}
\end{align}
where $x$ represents [Fe/H] and $y$ corresponds to $T_{\text{eff, norm}}$. The polynomial fitting coefficients and their corresponding root-mean-square errors (RMSE) are listed in Table~\ref{tab:2}. The RMSE values are generally consistent across all bands, around 0.24\,mag, except for the Mephisto $M_v$ band, which exhibits a slightly higher RMSE of 0.27\,mag in the validation set.

In Figs.~\ref{figure4}, \ref{figure5} and \ref{figure6}, we have also compared the polynomial fits with the observed data. Overall, the fits align well with the data. The residuals ($\Delta M$, defined as the difference between the polynomial fit and the binned medians) are also plotted as a function of stellar parameters. Within the well-sampled metallicity range ($-$0.5\,dex $<$ [Fe/H] $<$ 0.2\,dex) and temperature range (4500\,K $<$ $T_{\text{eff}}$ $<$ 5200\,K), the residuals remain small, typically within $\pm$0.05 to 0.1\,mag. However, larger residuals are observed at the extremes of metallicity and temperature.

\section{Discussion}\label{sec:4}

\subsection{Comparison with Previous Studies on RC Absolute Magnitudes} \label{sec:4.1}

Our empirical calibrations of RC absolute magnitudes in the Mephisto ($v$, $g$, $r$, $i$) and CSST ($g$, $r$, $i$) photometric systems reveal strong dependencies on stellar atmospheric parameters, particularly effective temperature and metallicity. These findings align with previous empirical and theoretical studies while extending our understanding through the use of a large, homogeneous, and carefully selected RC sample.

Earlier studies have consistently shown that RC stars are not perfect standard candles due to significant variations in their absolute magnitudes caused by metallicity and age—an effect known as population effects. For instance, \citet{pietrzynski2010araucaria} demonstrated that RC magnitudes in the optical $V$ and $I$ bands exhibit substantial scatter (up to 0.4\,mag) across different stellar populations. Their study of 23 nearby galaxies highlighted that neglecting population effects can lead to systematic errors in distance measurements. Similarly, \citet{ruiz2018empirical} used Gaia DR1 data to derive metallicity-dependent calibrations for RC absolute magnitudes, effective temperatures, and colors in optical-to-infrared bands. They found that RC absolute magnitudes vary with metallicity and color, with stronger variations observed at shorter wavelengths. Our results are consistent with these findings. 

Regarding effective temperature, our findings reveal a significant negative correlation between RC absolute magnitudes and $T_{\text{eff}}$, particularly in the Mephisto $v$ band, where magnitudes vary by approximately 2.0\,mag over 4500–5200\,K. This strong temperature dependence at shorter wavelengths was previously suggested by \citet{chen2017absolute}, who calibrated RC absolute magnitudes using Gaia DR1 and Kepler asteroseismic data and reported significant color (and thus temperature) dependencies. Additionally, our calibrations show minimal dependence on surface gravity, consistent with theoretical expectations and previous empirical results \citep[e.g.,][]{wan2015red,ruiz2018empirical}. By providing precise calibrations across the optical bands, our study enables accurate corrections for population effects, facilitating precise distance measurements and Galactic archaeology in upcoming Mephisto and CSST surveys.

\subsection{Validate Isochrone Predictions} \label{sec:4.2}

\begin{figure*}
    \centering
    \includegraphics[width=\textwidth]{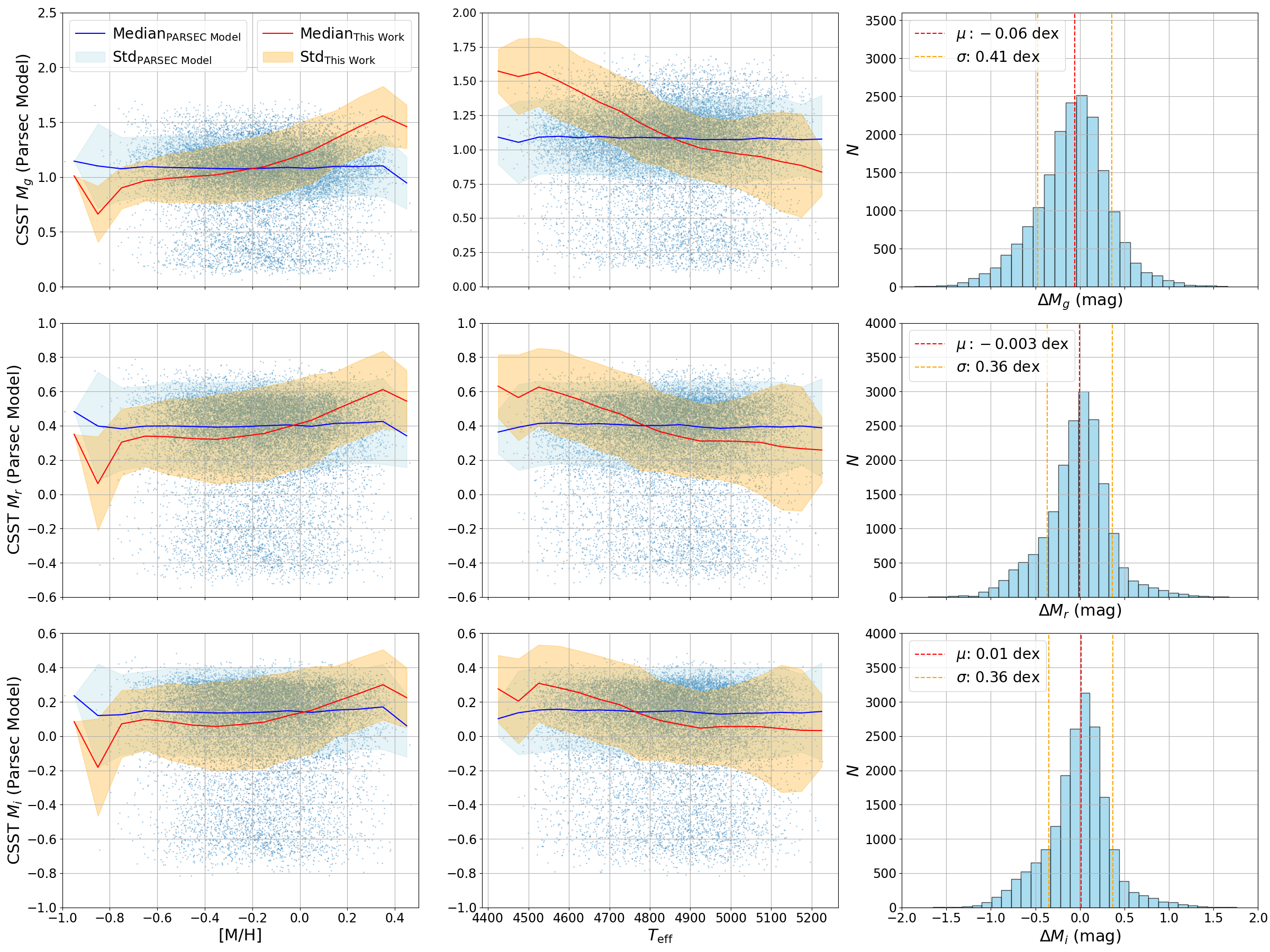} 
    \caption{Comparison of theoretical absolute magnitudes from PARSEC models with empirical results for RC stars in the CSST $g$, $r$, and $i$ bands. The left two columns show theoretical absolute magnitudes ($M_g, ~M_r, ~M_i$) as functions of [M/H], and $T_{\rm eff}$. Blue points represent individual stars. Blue solid lines and blue shaded regions indicate median trends and 1-$\sigma$ ranges from PARSEC models, respectively, while red solid lines and orange shaded regions denote median trends and 1-$\sigma$ ranges from our working data. The right column presents histograms of residuals ($\Delta M = M_{\text{This Work}} - M_{\text{PARSEC Model}}$) for each band, with the median and standard deviation displayed in the top-left corner of each panel.}
    \label{figure8}
\end{figure*}

Our empirical calibrations can be used to validate theoretical predictions by comparing the absolute magnitudes we derived for the CSST $g$, $r$, and $i$ bands with those predicted by the PARSEC isochrone models \citep{bressan2012parsec}. Using the PARSEC library, we generated theoretical CSST magnitudes for each RC star in our sample by interpolating their values based on their atmospheric parameters.  We note that we used [M/H] values from APOGEE DR17 here instead of [Fe/H] in this analysis rather than [Fe/H], as the PARSEC models provide only [M/H]. Fig.~\ref{figure8} illustrates the relationships between absolute magnitudes and stellar parameters for both our empirical results and theoretical predictions. The comparison reveals obvious differences in how parameter dependencies are captured by the two approaches.

In the CSST $g$ band, our empirical results show a clear increasing trend in absolute magnitudes with metallicity. In contrast, theoretical predictions show only a weak and slightly decreasing trend over the same range. Similarly, the dependence on $T_{\text{eff}}$ in the $g$ band is more pronounced in the empirical results, where magnitudes decrease significantly as $T_{\text{eff}}$ increases from 4500\,K to 5200\,K. This trend is not evident in theoretical predictions, which show minimal variation with $T_{\text{eff}}$. In the CSST $r$ and $i$ bands, both empirical and theoretical results exhibit weaker dependencies on metallicity and $T_{\text{eff}}$ than the $g$ band. However, even in these bands, empirical results display more pronounced trends than theoretical models.

The residuals $\Delta M = M_{\text{This Work}} - M_{\text{PARSEC Model}}$ in Fig.~\ref{figure8} further highlight these differences. In the CSST $g$ band, residuals exhibit significant scatter, with a median value of $\Delta M_g = -0.06$\,mag and a dispersion of 0.41\,mag. This suggests that theoretical predictions systematically deviate from empirical results and fail to capture observed dependencies on metallicity ([Fe/H]) and effective temperature ($T_{\text{eff}}$). In the CSST $r$ and $i$ bands, residuals are smaller, with median values of $\Delta M_r = -0.003$\,mag and $\Delta M_i = 0.01$\,mag, and dispersions of 0.36\,mag and 0.36\,mag, respectively. This indicates that theoretical models approximate RC absolute magnitudes in these bands more closely but still do not fully replicate the observed empirical trends.

Overall, these residual patterns confirm that while theoretical predictions provide a general approximation of RC absolute magnitudes, they fail to capture the full extent of parameter dependencies observed in our empirical calibrations. This underscores the importance of empirical calibrations for accurately characterizing RC stars, particularly for applications such as distance and metallicity determinations.

\subsection{Metallicity Estimation for RC Stars} \label{sec:3.3}

\begin{figure*}
    \centering
    \includegraphics[width=\textwidth]{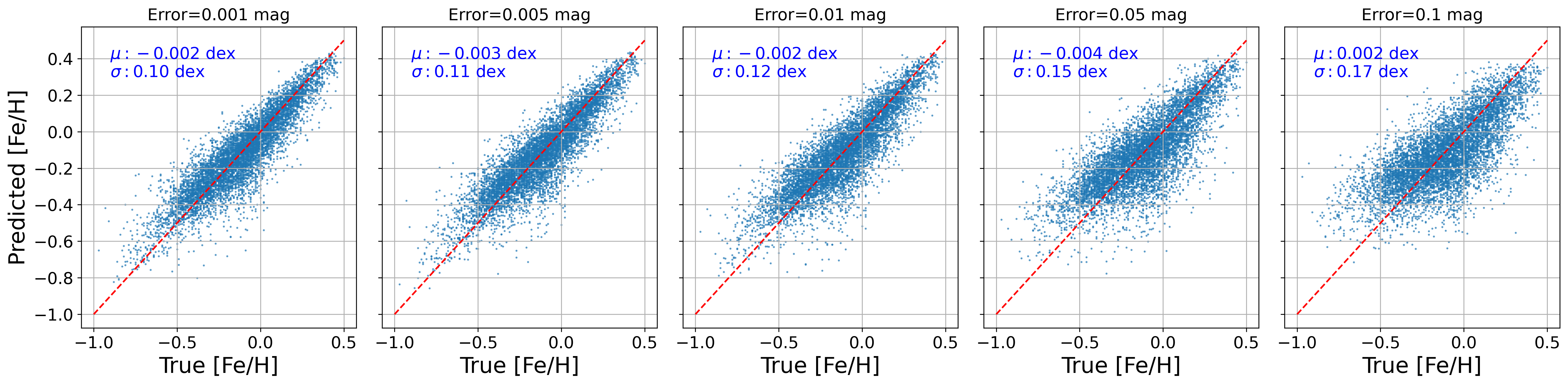}
    \includegraphics[width=\textwidth]{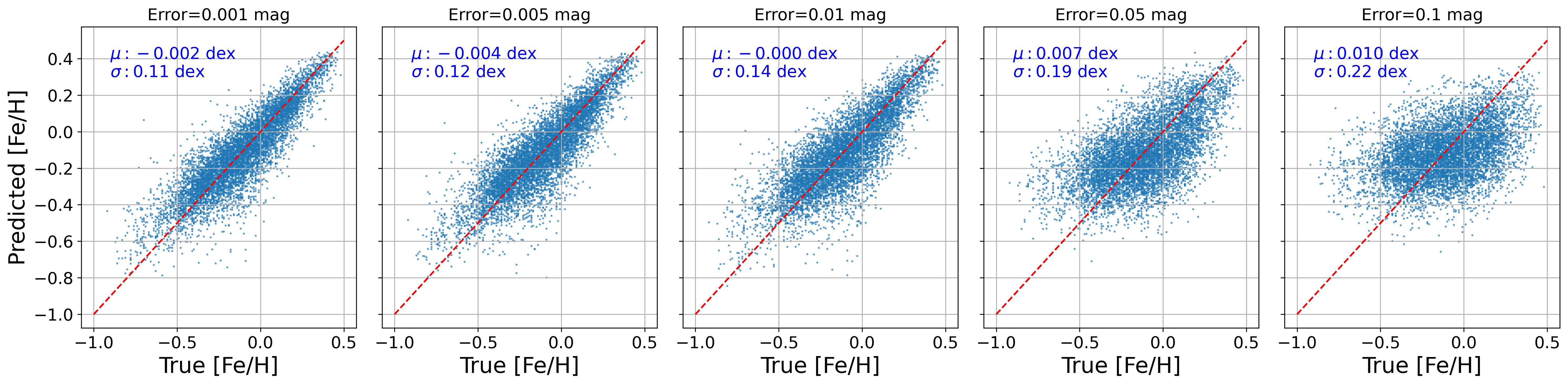}
    \caption{Comparison between predicted and true metallicities under different photometric error conditions. The upper panels correspond to the Mephisto filter system, while the lower panels show results from the CSST filter system. From left to right, photometric errors increase from 0.001 to 0.1\,mag. The red dashed lines indicate the one-to-one relation. The median offset and standard deviation for each case are displayed in the upper-left corner of each panel.}
    \label{figure7}
\end{figure*}

In the previous sections, we discussed the determination of absolute magnitudes for RC stars based on atmospheric parameters. Here, we explore the inverse problem—estimating stellar atmospheric parameters from Mephisto and CSST photometric data, with a focus on metallicity \([{\rm Fe/H}]\).

To achieve this, we employed the Random Forest (RF) algorithm, a robust ensemble learning method that combines predictions from multiple decision trees trained on different subsets of the data \citep{breiman2001random}. This approach effectively captures complex nonlinear relationships and provides stable predictions. Additionally, the RF algorithm ranks the importance of different input features, such as photometric colors, based on their contribution to the final predictions, offering insights into which colors are most informative for metallicity estimation. RF algorithms have been extensively applied to estimate parameters from photometric data \citep[e.g.,][]{Chen2019extin, sun2025photometric}, demonstrating good precision and robustness.

For the Mephisto system, we used three color indices: \((v - g)\), \((g - r)\), and \((r - i)\), while for the CSST system, we employed two color indices: \((g - r)\) and \((r - i)\). The RF regressor was implemented using the \texttt{scikit-learn} package \citep{pedregosa2011scikit}, with the number of estimators set to 100. Our RC sample was divided into training (70\%) and testing (30\%) subsets.

To assess the impact of observational uncertainties, we simulated five levels of photometric errors (0.001, 0.005, 0.01, 0.05, and 0.1\,mag) by adding Gaussian noise to the intrinsic colors. This allowed us to systematically evaluate how measurement uncertainties affect metallicity predictions. Fig.~\ref{figure7} presents a comparison between predicted and true metallicities for different photometric error levels.

For the Mephisto photometry, when photometric errors are small (\(\leq 0.01\)\,mag), the predicted metallicities closely match the true values, with standard deviations ranging from 0.10\,dex (for 0.001\,mag errors) to 0.12\,dex (for 0.01\,mag errors). The median offsets remain consistently below 0.003\,dex in absolute value. Even at higher photometric errors, the Mephisto system maintains relatively good precision, with standard deviations increasing to 0.15\,dex (for 0.05\,mag errors) and 0.17\,dex (for 0.1\,mag errors), while systematic offsets remain small (\(< 0.004\)\,dex in absolute value).

The CSST photometry exhibits larger scatter across all error levels. At small photometric errors (\(\leq 0.01\)\,mag), the standard deviation increases from 0.11\,dex (for 0.001\,mag errors) to 0.14\,dex (for 0.01\,mag errors), with minimal systematic offsets (\(-0.004\)\,dex to \(0\)\,dex). As photometric errors increase to 0.05\,mag and 0.1\,mag, both the scatter and systematic offsets become more pronounced: the standard deviations rise to 0.19\,dex and 0.22\,dex, while the systematic offsets increase to 0.007\,dex and 0.010\,dex, respectively.

Our results are consistent with \citet{huang2025stellar}, who applied the stellar locus method using Gaia DR3 XP synthetic photometry ($M_{BP - RP}$ and $M_{BP - G}$ colors) to estimate stellar metallicities, and achieved metallicity precisions of 0.05--0.10\,dex for both dwarfs and giants at [Fe/H] = 0 under typical photometric conditions. Similarly, we obtained precisions of 0.12\,dex and 0.14\,dex for the Mephisto and CSST photometric systems, respectively, under typical photometric uncertainties (\(\leq 0.01\)\,mag). This agreement also demonstrates that high-precision metallicity estimates can be achieved photometrically even using only two colors indices, such as the \((g - r)\) and \((r - i)\) colors we uesd for CSST.

For both systems, the accuracy of metallicity predictions declines more significantly for metal-poor stars (\([{\rm Fe/H}] < -0.5\)\,dex), as indicated by the increased scatter at lower metallicities in Fig.~\ref{figure7}. Nonetheless, when photometric errors are well controlled, both systems provide reliable metallicity estimates, with the Mephisto system consistently outperforming CSST. Our results suggest that achieving optimal metallicity precision requires photometric errors below 0.01\,mag, where both systems attain precision better than 0.14\,dex.

\section{Conclusion}\label{sec:5}

We have established precise empirical calibrations of RC absolute magnitudes in the Mephisto ($v, ~g, ~r, ~i$) and CSST ($g, ~r, ~i$) photometric systems, based on a carefully selected sample of 25,059 RC stars cross-matched between APOGEE and Gaia DR3 XP spectra. Our analysis reveals that RC absolute magnitudes exhibit significant dependencies on effective temperature and metallicity, with the strongest variations observed in the bluer bands.

In particular, the Mephisto $v$ band shows the highest sensitivity, with absolute magnitude variations reaching 2.0\,mag across the metallicity range ($-1.0\,\textrm{dex} < [\textrm{Fe/H}] < 0.5\,\textrm{dex}$) and the temperature range (4500--5200\,K). These dependencies weaken in redder bands, such as Mephisto $i$ and CSST $i$, where stellar population effects have a reduced influence. Our polynomial fitting method achieves high precision across all bands, ensuring reliable RC absolute magnitude and distance determinations. Photometric uncertainties contribute minimally to the total error, which is primarily dominated by distance uncertainties. While the Mephisto $v$ band exhibits slightly larger uncertainties due to increased photometric errors at shorter wavelengths, the overall calibrations remain robust for Galactic studies. Using a Random Forest regression approach, we evaluate the capability of Mephisto and CSST photometric colors to estimate metallicity. Under typical photometric conditions ($\leq 0.01$\,mag), our method achieves metallicity precisions of 0.12\,dex for Mephisto and 0.14\,dex for CSST. Even with larger photometric uncertainties (up to 0.05\,mag), both systems maintain reliable metallicity estimation performance, demonstrating the feasibility and robustness of photometric metallicity determinations for RC stars.

In summary, our empirical calibrations and metallicity estimation methods significantly enhance the scientific potential of future Mephisto and CSST photometric surveys. These calibrations will enable precise distance measurements and detailed chemical abundance studies across extensive regions of the Galactic disk, contributing to a deeper understanding of the structure, formation history, and chemical evolution of the Milky Way.

\normalem
\begin{acknowledgements}
This work is partially supported by the National Natural Science Foundation of China 12173034 and 12322304, the National Natural Science Foundation of Yunnan Province 202301AV070002 and the Xingdian talent support program of Yunnan Province. We acknowledge the science research grants from the China Manned Space Project with NO.\,CMS-CSST-2021-A09, CMS-CSST-2021-A08 and CMS-CSST-2021-B03. 

This publication makes use of data products from the European Space Agency (ESA) space mission Gaia. Gaia data are being processed by the Gaia Data Processing and Analysis Consortium (DPAC). Funding for the DPAC is provided by national institutions, in particular the institutions participating in the Gaia MultiLateral Agreement (MLA). The Gaia mission website is https://www.cosmos.esa.int/gaia. The Gaia archive website is https://archives.esac.esa.int/gaia.
\end{acknowledgements}
  
\bibliographystyle{raa}
\bibliography{bibtex}

\end{document}